\documentclass{optica-article}

\journal{opticajournal} 

\articletype{Research Article}

\usepackage{lineno}\usepackage{gensymb} 

\begin{document}

\title{Hybrid silicon-organic Huygens' metasurface for phase modulation}

\author{Sydney Mason,\authormark{1,*} and Ileana-Cristina Benea-Chelmus,\authormark{2,**}}

\address{\authormark{1}John A. Paulson School of Engineering and Applied Sciences, Harvard University, Cambridge, MA 02138, USA\\
\authormark{2}Hybrid Photonics Laboratory, École Polytechnique Fédérale de Lausanne, Lausanne, CH-1015, Switzerland}

\email{\authormark{*}sydneymason@college.harvard.edu} 
\email{\authormark{**}cristina.benea@epfl.ch}


\begin{abstract*} 
Spatial light modulators have desirable applications in sensing and free space communication because they create an interface between the optical and electronic realms. Electro-optic modulators allow for high-speed intensity manipulation of an electromagnetic wavefront. However, most surfaces of this sort pose limitations due to their ability to modulate intensity rather than phase. Here we investigate an electro-optic modulator formed from a silicon-organic Huygens' metasurface. In a simulation-based study, we discover a metasurface design immersed in high-performance electro-optic molecules that can achieve near-full resonant transmission with phase coverage over the full 2$\pi$ range. Through the electro-optic effect, we show 140\degree (0.79$\pi$) modulation over a range of -100 to 100 V at 1330 nm while maintaining near-constant transmitted field intensity (between 0.66 and 0.8). These results potentiate the fabrication of a high-speed spatial light modulator with the resolved parameters. 
\end{abstract*}

\section{Introduction}
With the capability to manipulate the amplitude, phase, and polarization of light\cite{Shaltout:06}, active photonics have penetrated different areas of research. Augmented reality (AR) and virtual reality (VR) systems need faster and more efficient sensors and delivery optics to keep up with the entertainment sector's rapid growth and increasing expectations. The optimization of free space optical communication requires devices that can engineer the properties of light at high speeds. Thin film metasurfaces allow us to control light intensity, polarization, and phase, serving as good candidates to address these applications\cite{MetasurfacesOverview:10, MetasurfacesOverview2:11}. Active surfaces of this sort have achieved dynamic wavefront manipulation by way of a variety of modulation formats, including but not limited to electro-optic\cite{Benea:08, EO_metasurfaces:13, EO_metasurfaces1:14}, thermo-optic\cite{TE_metasurfaces:17, TE_metasurfaces1:18}, nonlinear, and mechanical modulation\cite{h:20} and using phase change materials\cite{PC_metasurfaces:15, PC_metasurfaces1:16}. 

Achieving phase-dominated modulation is useful for beam steering and focusing, yet, compact and efficient phase modulation over the full 2$\pi$ range remains a difficult task. As a result, most active metasurface solutions target either intensity modulation or a compromise between the modulation of intensity and phase. Using integrated photonic circuits, the longer interaction length of waveguide-based modulation favors using effects like the Pockels effect for phase modulation\cite{n:21}. Given the short interaction length of sub-wavelength scale structures on a metasurface, it is necessary to rely on other effects for phase modulation. Metasurfaces can have resonant structures that simultaneously support intensity and phase\cite{e:22,l:23}, but, only in limited scenarios will these resonances support phase in isolation. Such scenarios require the matching of two different kinds of resonances to exploit the addition of the $\pi$ shift given by each resonance. These two resonances must be at the same wavelength and have the same linewidth, presenting increased design difficulty since this requires the balance of material losses and radiative losses\cite{BIC}. Previous work has optimized this balance using coupled plasmonic resonances and resonances supported by bound-states in the continuum (BIC), demonstrating 3$\pi$ phase modulation with constant reflection\cite{j:24} through the control of the carrier density of graphene. Other work uses permittivity tuning to achieve phase modulation of > 240$\degree$ in both transmission and reflection, although resonant dips reach near zero in both\cite{k:25}. Up to 300$\degree$ of phase modulation has been achieved using dual-gated reflective metasurfaces at low voltages exploiting the tuning of indium tin oxide as the active material, operating concurrently with an 89$\%$ intensity modulation\cite{l:23}. An alternative reflective metasurface solution reduces intensity variation to around 20$\%$ while achieving 270$\degree$ of electro-optic modulation\cite{d:26}. 

These solutions all operate in reflection, whereas phase modulation in transmission is also useful for applications like free-space optical communications and light detection and ranging (LiDAR). Transmissive Huygens’ metasurfaces create the potential for phase-dominated modulation through the overlapping of magnetic and electric dipole resonances. In aligning magnetic and electric dipole resonances with opposite polarizations they add to a constant total electric-magnetic field while the respective $\pi$ phase shift from each of the resonances combines to result in 2$\pi$ phase coverage\cite{DECKER:07}. Through control of the metasurface nanoantenna radii, the 2$\pi$ phase coverage has been harnessed to achieve beam deflection with transmission approaching unity\cite{g:27}. Dynamic beam steering up to 11$\degree$ has also been realized using the modulation of liquid crystal, with transmission dipping below 0.4\cite{f:28}, as well as low-voltage (1 V) beam steering at 15 ms switching speeds using a conductive polymer\cite{a:29}. An active Huygens’ surface that supports quasi-BIC resonances can modulate up to 240$\degree$ while maintaining transmission above 0.77\cite{m:30}. 

Although these demonstrations all show substantial progress in achieving phase modulation from transmissive Huygens’ metasurfaces, most operate at visible wavelengths or rely on relatively slow modulation effects. For communications applications, it is necessary to have high-speed optical modulation in the telecom wavelength region.  Here we propose and numerically validate an active Huygens’ metasurface that employs high-performing nonlinear organic molecules for phase modulation in the O-band, an ideal range for silicon photonics because there is no dispersion in fibers. Organic molecule JRD1 in polymethylmethacrylate (PMMA) (at a 50:50 concentration ratio) has a large nonlinear coefficient $r_{33}$ which when coupled with a metasurface can be exploited to achieve high-speed electro-optic modulation\cite{Benea:08,Benea2:09}. Electrical tuning has been demonstrated experimentally with JRD1 in a metasurface configuration at up to 3 GHz speeds with 3 dB bandwidth. JRD1 has high modulation efficiency and is easily fabricated with a metasurface due to deposition via spin-coating, making it an excellent candidate for modulation of the Huygens’ surface. JRD1 is also uniquely suited for our operation wavelength range, exhibiting the largest nonlinear effects in the O-band as compared to the typically used 1550 nm, with a higher refractive index (see Fig 1c.), higher $r_{33}$, and low losses.

\section{Theory and Methods}

\begin{figure}[!htbp]
\centering\includegraphics[width=9cm]{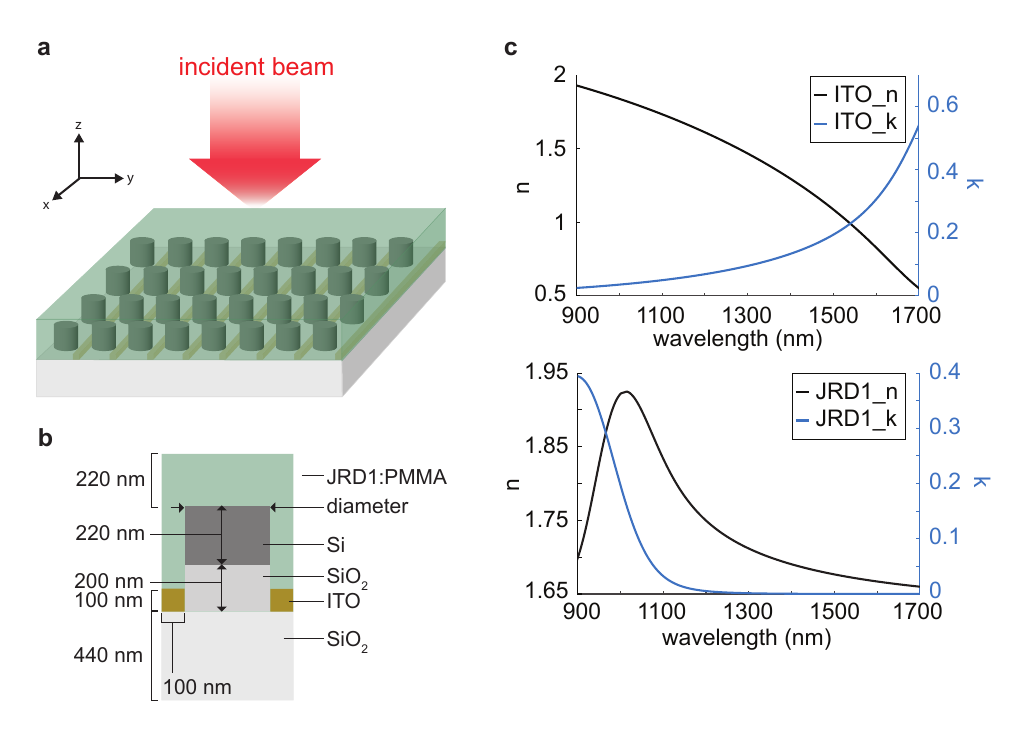}
\caption{a) The geometry of a metasurface-based electro-optic modulator with periodic Si nanopillars embedded in organic polymer with ITO electrodes. Si = silicon, JRD1:PMMA = JRD1 in polymethyl methacrylate at a 50:50 concentration, ITO = indium tin oxide, SiO$_2$ = silicon dioxide. b) Dimensions of a single metaatom cell of the metasurface with a silicon nanopillar height of 220 nm and a silicon dioxide pedestal height of  200 nm. c) Material properties (real (n) and imaginary (k) parts of the refractive indices) of JRD1:PMMA (50:50) and ITO.}
\label{fig:metasurfacegeo_matprops}
\end{figure}

The Huygens' metasurface, as shown in Figure~\ref{fig:metasurfacegeo_matprops}, consists of an array of periodic silicon (Si) nanopillars on a silica ({SiO\textsubscript{2}}) substrate. The pillars are embedded in a 640 nm thick layer of the JRD1:PMMA material and there are 100 nm wide and 100 nm tall interdigitated electrodes of indium tin oxide (ITO). ITO is a transparent, conductive material making it an ideal candidate for applying an electric field across the JRD1 while maintaining sufficient transmission. The refractive indices (real and imaginary) of both the dispersive materials, ITO and JRD1:PMMA, over the wavelength range of interest are shown in Figure~\ref{fig:metasurfacegeo_matprops}c)-d). We operate far from where $\epsilon$ = 0 for ITO, which is at around 1700 nm. At our operation wavelength, 1330 nm, n = 1.42 and k = 1.01 $\times$ $10^{-1}$ for ITO and n = 1.71 and k = 7.89 $\times$ $10^{-4}$ for JRD1. Additionally, in the surface design, the silicon pillar is raised from the SiO$_2$ substrate by a SiO$_2$ nano-pedestal. This pedestal adds extra distance between the silicon and the electrodes to reduce losses.

For a Huygens' surface, the relationship between the transmittance and reflection coefficients and resonance positions as a function of frequency can be described by Eq.\ref{eq:huygens_t} and Eq.\ref{eq:huygens_r}. In this case, $t$ and $r$ are the transmittance and reflection coeffficients respectively, $\gamma_m$ and $\gamma_e$ represent damping of the magnetic and electric fields and $\omega_m$ and $\omega_e$ are the resonance positions reflected by the Lorentzian dependence\cite{DECKER:07}. 
\begin{align}
    t = 1 + \frac{2i\gamma_e\omega}{\omega_e^2-\omega^2 - 2i\gamma_e\omega}+ \frac{2i\gamma_m\omega}{\omega_m^2-\omega^2 - 2i\gamma_m\omega} \label{eq:huygens_t} \\
    r = \frac{2i\gamma_e\omega}{\omega_e^2-\omega^2 - 2i\gamma_e\omega} -  \frac{2i\gamma_m\omega}{\omega_m^2-\omega^2 - 2i\gamma_m\omega}
    \label{eq:huygens_r}
\end{align}

To achieve optical control we manipulate the resonant structures by changing the refractive index of the resonance-supporting surface using the Pockels effect. After poling the organic molecules, by applying a time-dependent voltage $V_{rf}(t)$, the refractive index can be altered to follow this time dependency, yielding a refractive index change $\Delta n(t) = -\frac{1}{2}n^{3}r_{33}E_{rf}(t)$, where $E_{rf}(t)$ is the built-in time-dependent electric field due to the applied voltage $V_{rf}(t)$ and $n$ is the initial refractive index of the material \cite{Benea2:09}. The $r_{33}$ of JRD1 can be engineered to be 100 pm/V\cite{Benea2:09} at 1550 nm and experimental values of the organic molecule at 1310 nm has been shown to achieve an $r_{33}$ of 343 pm/V\cite{JRD1_r33}. Considering this, we can expect an $r_{33}$ significantly greater than 100 pm/V at 1330 nm. Using Eq.\ref{eq:huygens_t} and assuming the most simple case of overlapping magnetic and electric resonances ($\omega_e$=$\omega_m=\omega_{res}$, $\gamma_e$=$\gamma_m=\gamma$), we can represent the phase ($\phi$) of the transmitted field as:
\begin{align}
    \phi = atan\frac{4\gamma\omega(\omega_{res}^2-\omega^2)}{(\omega_{res}^2-\omega^2)^2 - \gamma^2\omega^2}
\label{eq:phiandomega}
\end{align}

Additionally, we can represent the resonant shift relative to an applied voltage across the electrodes:
\begin{align}
     \frac{\Delta\omega_{res}}{\Delta V} = \frac{\omega_{res}\Gamma n^2 r_{33}}{2L}
\label{eq:omegaandref}
\end{align}

Where $\Gamma$ is the overlapping factor between the organic material's $r_{33}$, the optical field, and the RF field, and $L$ is the separation between the electrodes. Using Eq. \ref{eq:phiandomega}, we evaluate $\frac{\Delta\phi}{\Delta\omega_{res}}$ at $\omega = \omega_{res}$ and then multiply it with Eq.\ref{eq:omegaandref} to model the total electro-optic phase shift from our hybrid-organic Huygens' surface as:
\begin{align}
     \frac{\Delta\phi}{\Delta V} 
     =\frac{-4\omega_{res}\Gamma n^2 r_{33}}{\gamma L}
\label{eq:phiandvoltage}
\end{align}

\begin{figure}[!htbp]
\centering\includegraphics[width=9cm]{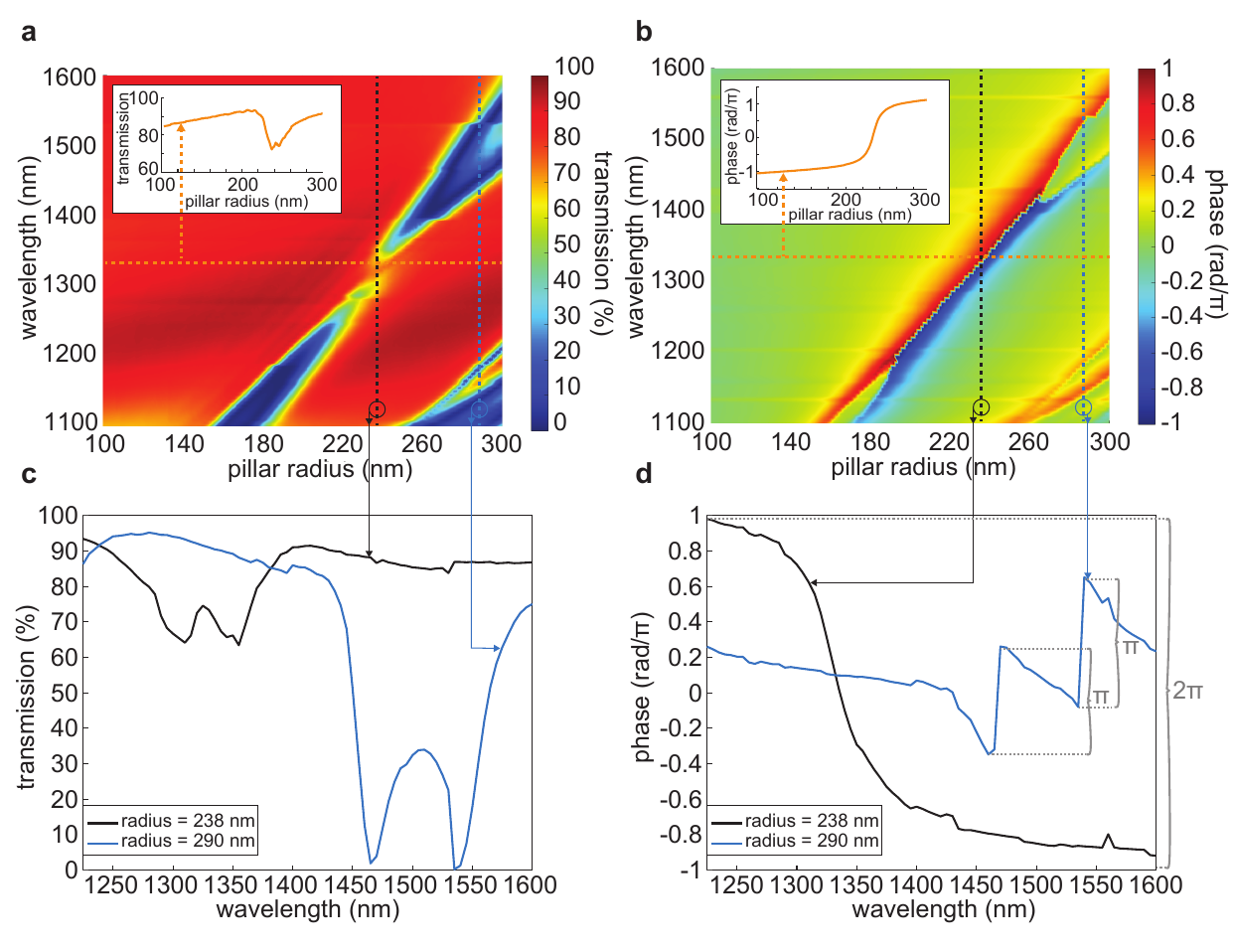}
\caption{a) Transmission of light through a metasurface with periodic nanopillars of height 220 nm and varying radii. The nanopillars are embedded in a film of organic molecules JRD1, set between interdigitated electrodes of ITO. The surface rests on a silica substrate. The pillar radius with near-constant transmission across the wavelength range is denoted with the black dotted line. The case where the magnetic and electric resonances are separated is denoted with the blue dotted line. Additionally, the inset shows the near-constant transmission for the range of pillar radii between 100 and 300 nm at 1330 nm of incident light. b) Phase of light through the same metasurface described in a), the pillar radius with 2$\pi$ phase coverage (the same as the near-constant transmission in a)) is denoted with the black dotted line and corresponding case from a) with the blue dotted line. The inset shows an alternative approach, a 2$\pi$ phase coverage that can also be achieved at a constant wavelength (1330 nm) while varying the pillar radii. c) Transmission spectra of two radii, where the Huygens' condition is (r = 238 nm) and is not achieved (r = 290 nm). d) Phase of the same two radii as c)}
\label{fig:colormaps}
\end{figure}

Given a preliminary estimate of $r_{33}$ at 200 pm/V, 200 V of applied voltage, Lorentzian linewidth ($\gamma$) of 50 nm, overlap factor $\Gamma$ of 0.1 and $\lambda_{res}$ of 1330 nm, we predict phase modulation of approximately 0.73$\pi$, using Eq.\ref{eq:phiandvoltage}. Electrodynamical software Lumerical FDTD (finite-difference time-domain) is used to perform a sweep of design parameters to target the nanopillar geometry that can achieve the Huygens' surface criteria. The design parameter of particular interest is the pillar diameter while the pillar height (220 nm) is kept constant. The incident light source wavelength is also swept in the telecom range between 1100 and 1600 nm with 5 nm steps. The results of these simulations are shown in Figure~\ref{fig:colormaps}a)-b). The observed region of interest is discovered to be a pillar radius between 220 nm and 250 nm, where the magnetic and electric resonances converge and the transmission remains above a 50$\%$ threshold. For this paper, we define the transmission to be the ratio of ${|E_{out}|}$/${|E_{in}|}$, where $E_{out}$ is the electric field measured after the metasurface and $E_{in}$ is the electric field of the incident light. Within this region we observe a 2$\pi$ phase shift, as can be seen in Figure~\ref{fig:colormaps}b). Following this characterization, further simulations are conducted with increased granularity to find the optimal parameters. 

\section{Simulation Results}
\begin{figure}[!htbp]
\centering\includegraphics[width=9cm]{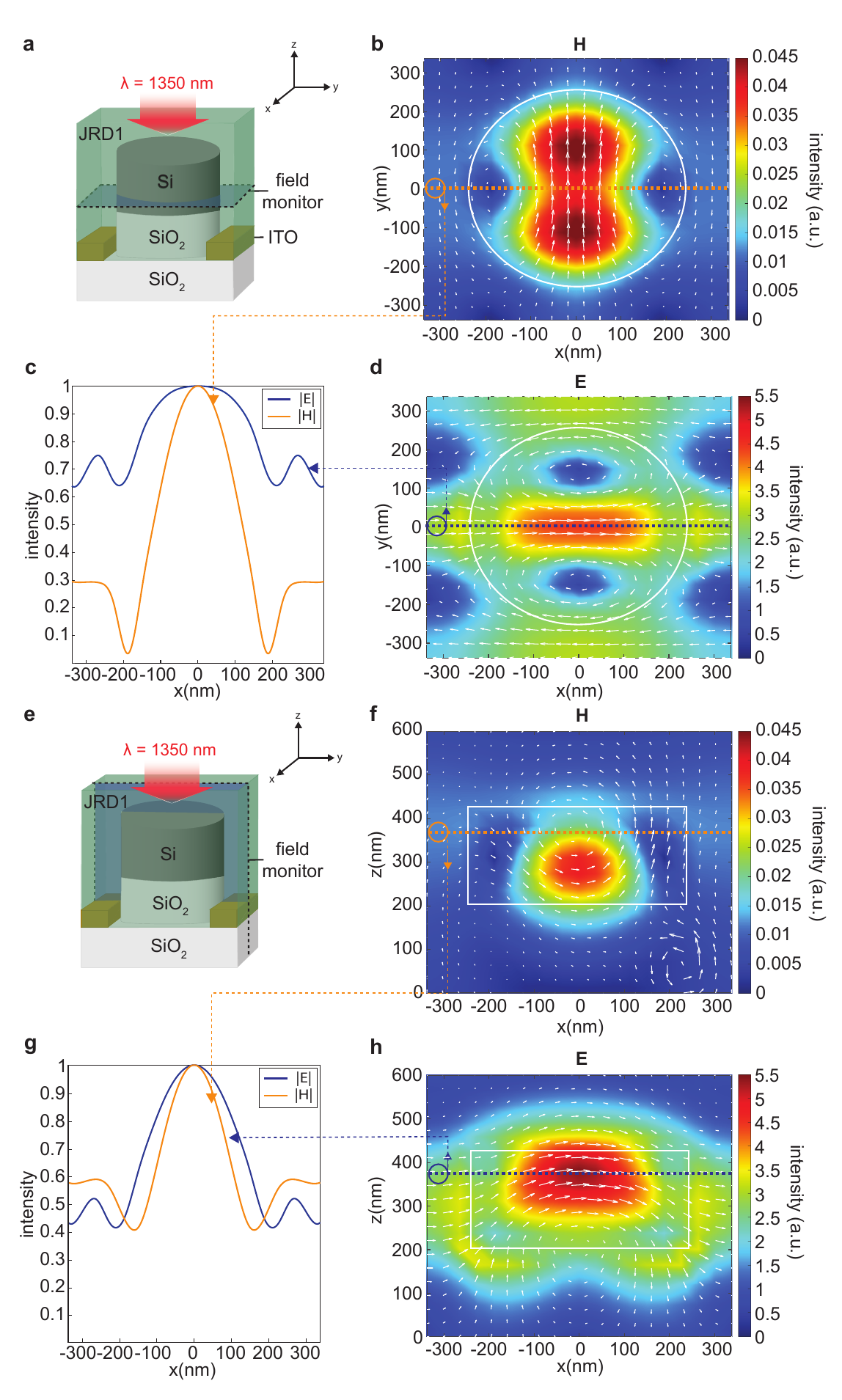}
\caption{Field profile analysis of a silicon nanopillar embedded in JRD1 on glass a) Simulation field monitor orientation used for b)-d), b) Magnetic field profile in z-normal plane of nanopillar with 238 nm radius at 1350 nm incident light, c) Electric and magnetic field magnitudes at y = 0 of b) and d), d) Electric field profile in z-normal plane of nanopillar with 238 nm radius at 1350 nm incident light, e) Simulation field monitor orientation used for f)-h), f) Magnetic field profile in y-normal plane of nanopillar with 238 nm radius at 1350 nm incident light, c) Electric and magnetic field magnitudes at z = 300 nm of b) and d), d) Electric field profile in y-normal plane of nanopillar with 238 nm radius at 1350 nm incident light}
\label{fig:fieldprofs}
\end{figure}

As a result of this simulation-based study, we discover a Huygens' surface with nanopillar height of 220 nm and radius of 238 nm. Over an incident wavelength range of 1100 to 1600 nm we observe 2$\pi$ phase coverage while maintaining transmission above 0.65. Through simulation, we can clearly denote the conditions where the electric and magnetic fields are aligned within the nanoparticle and where they are not, as described by the transmission and phase plots in Figure~\ref{fig:colormaps}c)-d). With a nanopillar radii of 290 nm, the expected resonant dips occur from the electric and magnetic dipoles at different wavelengths. This results in two separate $\pi$ phase shifts at the separate wavelength positions, as opposed to the 2$\pi$ shift and near-constant transmission with a radius of 238 nm. In the case where this alignment is achieved, the resulting field profiles within a single metaatom cell are shown in Figure~\ref{fig:fieldprofs}.

\begin{figure}[!htbp]
\centering\includegraphics[width=9cm]{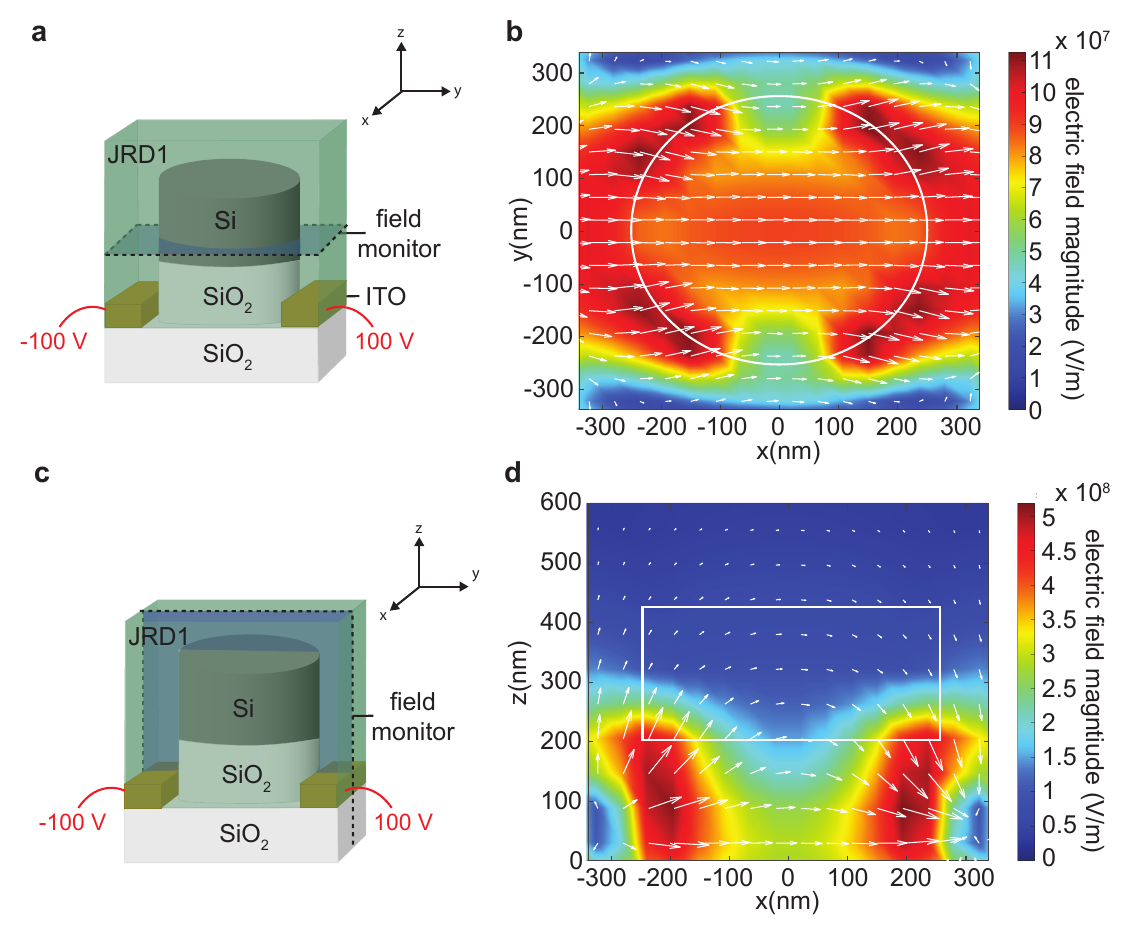}
\caption{a) Electric field monitor position (in the z-normal plane at z = 310 nm) during simulation of a DC field in a single metaatom cell. -100 V is applied to one electrode while 100 V is applied to to other, creating the electric field across the metaatom. b) Electric field distribution in the z-normal plane (as indicated by the axes in a)). The white arrows illustrate the vectorial components of the electric field and the white circle indicates the silicon pillar position. c) Electric field monitor position during simulation of a DC field in a single metaatom cell, monitoring in the y-normal plane (at y = 0). The -100 V and 100 V are applied to the electrodes as in a). d) Electric field distribution in the y-normal plane. As in b) the arrows represent the vectorial components of the field and the white box the silicon pillar.}
\label{fig:DCfieldprofs}
\end{figure}

The electro-optic modulation potential of our Huygens' metasurface design is validated through the simulation of an applied electric field across the JRD1 material. In order to achieve meaningful phase modulation, it is necessary to have sufficient spatial overlap and vectorial alignment of the optical field and the applied RF field to efficiently exploit the $r_{33}$ of the organic material . The polarization of the organic molecules is conveniently aligned with the RF field direction through the poling of the material with the same electrodes that apply the RF field. The electric field distribution in a single metaatom cell with an applied DC field (-100 V on one electrode, 100 V on the other) is indicated in Figure \ref{fig:DCfieldprofs}. When this field distribution is compared to the optical field in Figure \ref{fig:fieldprofs}, we can see that the electric component of the optical field in the organic material in area directly surrounding the nanopillar aligns considerably well with the same region in the RF distribution where the field is strongest. 

\begin{figure}[!htbp]
\centering\includegraphics[width=9cm]{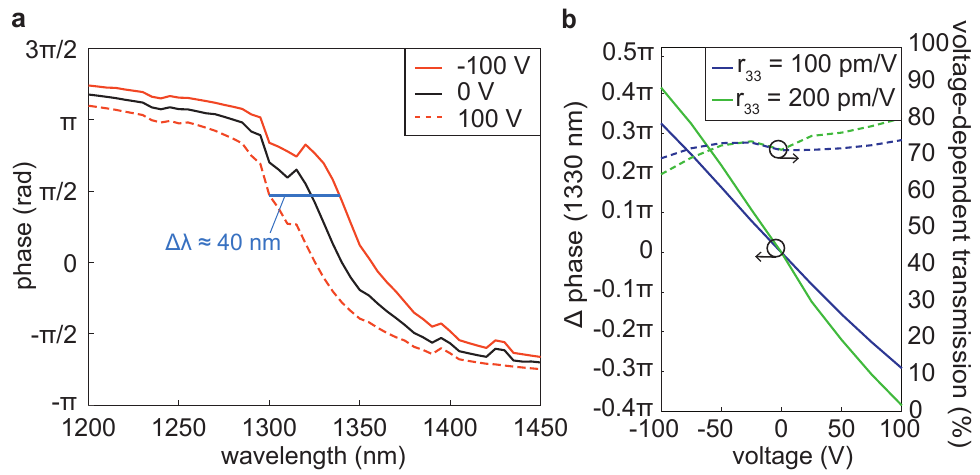}
\caption{a) The phase shift of transmitted radiation through a Huygens' metasurface with - 100 and 100 V of applied voltage. The ITO electrodes create an electric field that is aligned with the organic material's $r_{33}$ polarization, altering the material's refractive index and creating a 40 nm shift in the phase. b) Phase modulation relative to the applied voltage at 1330 nm for an $r_{33}$ of both 100 pm/V and 200 pm/V. The modulation is dominated in the phase (left axis), while the transmission during modulation remains near-constant (right axis).}
\label{fig:operation}
\end{figure}

Using $\Delta n(t) = -\frac{1}{2}n^{3}r_{33}E_{rf}(t)$ as an analytical model for the corresponding change in refractive index with applied voltage, we simulate an applied voltage ranging from -100 to 100 V and record the transmitted field's phase frequency shift (Figure \ref{fig:operation}). We simulate a conservative value for $r_{33}$ from the characterization at 1550 nm (100 pm/V) as well as an $r_{33}$  of 200 pm/V because we assume the $r_{33}$ to be much higher at 1330 nm. In the case of $r_{33}$ = 200 pm/V, we observe a wavelength shift of approximately 40 nm of the phase curve, corresponding to a phase shift of 140\degree ($\approx$0.8$\pi$) with an applied voltage from - 100 to 100 V (indicated in Figure~\ref{fig:operation}a)). Here we define the $\Delta$ phase as the difference between $\phi$ at 0 V and $\phi$ with an applied voltage across the electrodes. 

\section{Discussion}
The purpose of this study was to explore whether an electro-optic phase modulator made from a Huygens' metasurface is a viable application worth investigating experimentally. The high efficiency of JRD1 coupled with our Huygens' metasurface both operating in the O-band, provides a feasible platform for phase modulation for optical communications.  Fabrication processes for metasurfaces of this sort have been prevalent in the field since the 2010s\cite{MSfab:19}. Therefore, the potential fabrication of this surface for experimental realization would follow similar well-known nanofabrication protocols, easily integrating the deposition of JRD1 through spin-coating.  Harnessing the tunability of metasurfaces and 2$\pi$ coverage given by varying pillar diameters, the design can be further optimized to create a predefined tilt of the transmitted light by introducing a built-in linear phase gradient. In principle, magnetic and electric dipole alignment should produce constant transmission across the frequency region of interest. Yet, we recognize that this ideal condition is very difficult to achieve in actuality. In our results, we observe small dips in the transmission curves in the 1275 - 1375 nm wavelength range due to the slight mismatch of magnetic and electric resonant widths. In future research, the losses of the different materials could be further optimized to achieve a more constant transmission. 

\section{Conclusion}
In adding dynamic wavefront manipulation to a Huygens' metasurface by way of our proposed electro-optic phase modulation, we can achieve phase modulation of $\Delta 0.8\pi$ at 1330 nm with limited change in transmitted intensity. This research suggests the use of a novel active Huygens' metasurface platform for efficient electro-optic phase modulation in the O-band. Free-space modulators with 2$\pi$ phase coverage and constant transmission are desirable for applications in the free-space domain in cameras, for telecommunications applications, and for sensing in autonomous vehicles.

\newpage
\begin{backmatter}
\bmsection{Funding} S.M. and I.-C.B.-C. acknowledge funding from the Swiss National Science Foundation under PRIMA Grant No. 201547.
\bmsection{Acknowledgments} S.M. acknowledges I.-C.B.-C. for their mentorship and enlightening discussions. I.-C.B.-C. inspired this study and S.M. and I.-C.B.-C. conceptualized the idea. S.M. conducted the computation and numerical simulations and wrote the paper.

\bmsection{Disclosures} The authors declare no conflicts of interest.
\bmsection{Data Availability Statement} The data collected will be made available on the Zenodo database prior to publication.
\end{backmatter}


\bibliography{sample}

\end{document}